\documentclass[11pt, a4paper]{article}
\pdfoutput=1

\usepackage{jheppub}

\usepackage{subcaption} 

\usepackage[utf8]{inputenc} 
\usepackage{physics} 
\usepackage{booktabs} 
\usepackage{multirow} 
\usepackage[dvipsnames]{xcolor}

\usepackage{tikz}
\usetikzlibrary{arrows,shapes}
\usetikzlibrary{trees}
\usetikzlibrary{matrix,arrows} 				
\usetikzlibrary{positioning}				
\usetikzlibrary{calc,through}				
\usetikzlibrary{decorations.pathreplacing}  
\usepackage{pgffor}							
\usetikzlibrary{decorations.pathmorphing}	
\usetikzlibrary{decorations.markings}
\tikzset{
	vector/.style={decorate, decoration={snake}, draw},
	provector/.style={decorate, decoration={snake,amplitude=2.5pt}, draw},
	antivector/.style={decorate, decoration={snake,amplitude=-2.5pt}, draw},
	fermion/.style={draw=black, postaction={decorate},
	decoration={markings,mark=at position .55 with {\arrow[draw=black]{>}}}},
	fermionbar/.style={draw=black, postaction={decorate},
	decoration={markings,mark=at position .55 with {\arrow[draw=black]{<}}}},
	fermionnoarrow/.style={draw=black},
	gluon/.style={decorate, draw=black,
	decoration={coil,amplitude=4pt, segment length=5pt}},
	scalar/.style={dashed,draw=black, postaction={decorate},
	decoration={markings,mark=at position .55 with {\arrow[draw=black]{>}}}},
	scalarbar/.style={dashed,draw=black, postaction={decorate},
	decoration={markings,mark=at position .55 with {\arrow[draw=black]{<}}}},
	scalarnoarrow/.style={dashed,draw=black},
	electron/.style={draw=black, postaction={decorate},
	decoration={markings,mark=at position .55 with {\arrow[draw=black]{>}}}},
	bigvector/.style={decorate, decoration={snake,amplitude=4pt}, draw},
}

\newcommand{\arxiv}[1]{arXiv:\href{http://www.arxiv.org/abs/#1}{#1}}
\newcommand{\doi}[2]{\href{http://doi.org/#1}{#2}}

\def\Da{\Delta_a}

\def\dm{\Delta m^2}

\def\dmt{\Delta \tilde m^2}
\def\dmtbar{\Delta \tilde {\bar m}^2}

\newcommand{\asym}[1]{\mathcal{A}^\mathrm{#1}}
\newcommand{\barasym}[1]{\mathcal{\bar A}^\mathrm{#1}}
\newcommand{\comp}[2]{A^{\mathrm{#1};\mathrm{#2}}}

\title{Do T asymmetries for neutrino oscillations in uniform matter have a CP-even component?}

\author{Jos\'e Bernab\'eu}
\author{and Alejandro Segarra}
\affiliation{Departament de F\'isica Te\`orica and IFIC, Universitat de Val\`encia - CSIC, E-46100, Spain}

\emailAdd{Jose.Bernabeu@uv.es}
\emailAdd{Alejandro.Segarra@uv.es}

\abstract{
	Observables of neutrino oscillations in matter have, in general, 
	contributions from the effective matter potential. 
	It contaminates the CP violation asymmetry adding a fake effect
	that has been recently disentangled from the genuine one
	by their different behavior under T and CPT.
	Is the genuine T-odd CPT-invariant component of the CP asymmetry 
	coincident with the T asymmetry? 
	Contrary to CP, matter effects in uniform matter cannot 
	induce by themselves a non-vanishing T asymmetry;
	however, the question of the title remained open. 
	We demonstrate that, in the presence of genuine CP violation, 
	there is a new non-vanishing CP-even, and so CPT-odd, component in the T asymmetry in matter, 
	which is of odd-parity in both the phase $\delta$ of the flavor mixing and the matter parameter $a$. 
	The two disentangled components, 
	genuine $\comp{T}{CP}_{\alpha\beta}$
	and fake $\comp{T}{CPT}_{\alpha\beta}$, 
	could be experimentally separated by the measurement of the two T asymmetries in matter 
	($\nu_\alpha \leftrightarrow \nu_\beta$) and ($\bar\nu_\alpha \leftrightarrow \bar\nu_\beta$). 
	For the ($\nu_\mu \leftrightarrow \nu_e$) transitions, the energy
	dependence of the new $\comp{T}{CPT}_{\mu e}$ component is like the 
	matter-induced term $\comp{CP}{CPT}_{\mu e}$ of the CP asymmetry
	which is odd under a change of the neutrino mass hierarchy.
	We have thus completed the physics involved in all observable asymmetries in matter 
	by means of their disentanglement into the three independent components, 
	genuine $\comp{CP}{T}_{\alpha\beta}$ 
	and fake $\comp{CP}{CPT}_{\alpha\beta}$ and $\comp{T}{CPT}_{\alpha\beta}$.
}

\begin{document}

\maketitle
\flushbottom

\section{Introduction}
\label{sec:intro}

Neutrino oscillations from a terrestrial accelerator source 
take place through their propagation in matter of the Earth mantle. 
There are matter effects originated in the interference between
the forward scattering amplitude of electron neutrinos with matter electrons
and the free propagation.
Matter being CP ---and CPT--- asymmetric leads to 
well known fake contributions to the CP asymmetry 
$\asym{CP}_{\alpha\beta}$ 
between the $\alpha\to\beta$ flavor transition probabilities for neutrinos and antineutrinos. 
Recently, the disentanglement of genuine and matter-induced CP violation has been solved 
by means of a theorem~\cite{disentanglingPRL} able to separate the experimental CP asymmetry 
into two components with well defined behavior under the discrete symmetries: 
a genuine term $\comp{CP}{T}_{\alpha\beta}$ odd under T and CPT-invariant,
and a fake term $\comp{CP}{CPT}_{\alpha\beta}$ odd under CPT and T-invariant. 
In addition, their peculiar different energy distributions at a fixed baseline 
provide signatures~\cite{disentanglingJHEP} for their separation in an actual experiment.
Their definite different parities under the baseline, 
the matter potential and the genuine CPV phase allow 
the use of guiding simple and precise enough expressions in terms of vacuum parameters.

A question immediately arises: 
are these $\comp{CP}{T}_{\alpha\beta}$ and $\comp{CP}{CPT}_{\alpha\beta}$
components coincident with the T and CPT asymmetries, respectively?
As the genuine T-odd $\comp{CP}{T}_{\alpha\beta}$ component is CPT-invariant and thus CP-odd, 
a positive answer to this question would be equivalent to claim 
the absence of fake matter-induced terms in the T asymmetries, 
and hence a negative answer to the title of this paper. 
In the literature on T asymmetries for neutrino oscillations~\cite{T1,T2,T3,T4,T5,scott,xing,T6,freund,T7,T8,T9,T10}
one does not find definite claims on this question,
even for a T-symmetric matter between the source and the detector
(and, a fortiori, for uniform matter) 
neither in one nor the other sense of the response. 
Contrary to the fake $\comp{CP}{CPT}_{\alpha\beta}$ component of the CP asymmetry, 
which exists even in the absence of true CP violation, 
if the medium is T-symmetric it, \emph{by itself}, 
cannot generate a T asymmetry in neutrino oscillations.
However, in the presence of genuine CP violation,
this last reasoning does not lead to a definite conclusion 
whether the entire T asymmetry is also CP-odd and CPT-invariant. 
In this last case,
the genuine $\comp{CP}{T}_{\alpha\beta}$ component in the disentangled CP asymmetry of
Ref.~\cite{disentanglingPRL} could be separately measured by the T asymmetry. 
On the contrary, a positive answer to the title of this paper
would mean that the medium generates an additional $\comp{T}{CPT}_{\alpha\beta}$
component which is CP-even by the combined effect of genuine \emph{and} matter amplitudes.
This CPT-odd component would then be a fake effect even for a T-symmetric medium.

We plan to solve this open question in this paper and,
anticipating a possible presence of $\comp{T}{CPT}_{\alpha\beta}$, 
prove an analogous T asymmetries Disentanglement Theorem into two components: 
a CP-odd CPT-invariant $\comp{T}{CP}_{\alpha\beta}$ component which is identified 
with the known genuine $\comp{CP}{T}_{\alpha\beta}$ 
and a new CPT-odd CP-invariant $\comp{T}{CPT}_{\alpha\beta}$ component. 
As a lemma for the fake CPT asymmetries, 
written as a combination of CP and T asymmetries, 
they would be disentangled into the two components
$\comp{CPT}{CP}_{\alpha\beta} = \comp{CP}{CPT}_{\alpha\beta}$
of the CP asymmetry
and $\comp{CPT}{T}_{\alpha\beta} = \comp{T}{CPT}_{\alpha\beta}$
of the T asymmetry.
No experimental asymmetry would be genuine by itself:
only the $\comp{CP}{T}_{\alpha\beta} = \comp{T}{CP}_{\alpha\beta}$ component is so,
whereas $\comp{CP}{CPT}_{\alpha\beta} = \comp{CPT}{CP}_{\alpha\beta}$ is purely a medium effect
and $\comp{T}{CPT}_{\alpha\beta} = \comp{CPT}{T}_{\alpha\beta}$ combines genuine and matter-induced amplitudes.

The paper is organized as follows. 
In Section~\ref{sec:Tdisentangling} we prove the second Disentanglement Theorem, 
this time for the T asymmetries in matter, 
and identify the relevant rephasing-invariant mixings and oscillation factors 
appearing separately in each of the two components
$\comp{T}{CPT}_{\alpha\beta}$ and $\comp{T}{CP}_{\alpha\beta}$,
which are both odd functions of the baseline $L$. 
For the three-family Hamiltonian in terms of 
vacuum parameters and the matter potential, 
we establish in Section~\ref{sec:Tanalytic} the definite parities of the two components under
the phase $\delta$ of the flavor mixing matrix and the matter parameter $a$. 
We calculate their peculiar energy dependencies
in the energy region between the two MSW~\cite{MSW-W,MSW-MS} resonances,
both numerically and using the analytic perturbation expansion performed in Ref.~\cite{disentanglingJHEP},
for the golden $\nu_\mu \to \nu_e$ channel.
%
Section~\ref{sec:CPT} discusses the specific combinations which appear in CPT asymmetries,
as a lemma of the two Disentanglement Theorems.
The main conclusions are given in Section~\ref{sec:conclusions}.
%

\section{T asymmetry disentanglement theorem}
\label{sec:Tdisentangling}

In this Section we study the time evolution of neutrinos propagating in matter
in terms of their effective masses $\tilde m$ and mixings $\tilde U$.
Notice that both ingredients are energy dependent.
The flavor oscillation probability is
\begin{equation}
	\label{eq:Pab}
	P(\nu_\alpha \to \nu_\beta)
	= \delta_{\alpha\beta}
	-4\sum_{j<i}\mathrm{Re}~\tilde J_{\alpha\beta}^{ij}\,
	\sin^2 \tilde \Delta_{ij}\,
	-2\sum_{j<i}\mathrm{Im}~\tilde J_{\alpha\beta}^{ij}\,
	\sin 2\tilde \Delta_{ij}\,,
\end{equation}
where 
$\tilde J_{\alpha\beta}^{ij} \equiv 
\tilde U_{\alpha i} \tilde U^*_{\alpha j}
\tilde U^*_{\beta i} \tilde U_{\beta j}$
are the rephasing-invariant mixings
and
$\tilde \Delta_{ij} \equiv \frac{\Delta\tilde m^2_{ij} L}{4 E}$
the oscillation phases in matter.
In terms of these quantities, the disentangled components of the CP asymmetry,
$\asym{CP}_{\alpha\beta} \equiv 
P_{\alpha\beta} - \bar P_{\alpha\beta}
= \comp{CP}{T}_{\alpha\beta} + \comp{CP}{CPT}_{\alpha\beta}$,
are~\cite{disentanglingPRL}
\begin{subequations}
	\label{eqs:compsACP}
	 \begin{align}
		\label{eq:ACP(T)}
	    \comp{CP}{T}_{\alpha\beta} &=
	    -2\sum_{j<i} \left[
	   	\mathrm{Im}~\tilde J_{\alpha\beta}^{ij}\,\sin 2\tilde \Delta_{ij}
		-\mathrm{Im}~\tilde {\bar J}_{\alpha\beta}^{ij}\,\sin 2\tilde {\bar \Delta}_{ij}
		\right]\,,\\
		\label{eq:ACP(CPT)}
	    \comp{CP}{CPT}_{\alpha\beta} &= 
	    -4\sum_{j<i} \left[
	   	\mathrm{Re}~\tilde J_{\alpha\beta}^{ij}\,\sin^2 \tilde \Delta_{ij}
		-\mathrm{Re}~\tilde {\bar J}_{\alpha\beta}^{ij}\,\sin^2 \tilde {\bar \Delta}_{ij}
		\right]\,.
	 \end{align}
\end{subequations}

The definition of the rephasing-invariant mixings says that 
$\tilde J_{\beta\alpha}^{ij} = (\tilde J_{\alpha\beta}^{ij})^*$,
so the $\alpha \leftrightarrow \beta$ exchange in the T asymmetry leads to
\begin{equation}
	\label{eq:asymT}
	\asym{T}_{\alpha\beta} \equiv P_{\alpha\beta}-P_{\beta\alpha} =
	-4\sum_{j<i}\mathrm{Im}~\tilde J_{\alpha\beta}^{ij}\,
	\sin 2\tilde \Delta_{ij}\,,
\end{equation}
which is an odd function of the baseline $L$, as imposed by its T-odd character.
Notice that, even though matter may affect the T asymmetry,
it cannot generate a non-vanishing T asymmetry in the absence of genuine
T violation ---the real mixing matrix would ensure a vanishing $\asym{T}_{\alpha\beta}$.
Therefore, contrary to the CP asymmetry case,
a non-vanishing T asymmetry in matter is a proof of
the existence of genuine CP violation in the lepton sector.
Strangely enough, this reasoning does not prove 
that CP violation has actually been seen in the T asymmetry.

To check whether this quantity violates CP or CPT, as well as T,
it should be compared with its $\bar \nu$ equivalent,
\begin{equation}
	\label{eq:asymTbar}
	\barasym{T}_{\alpha\beta} \equiv \bar P_{\alpha\beta}- \bar P_{\beta\alpha} =
	-4\sum_{j<i}\mathrm{Im}~\tilde {\bar J}_{\alpha\beta}^{ij}\,
	\sin 2\tilde {\bar \Delta}_{ij}\,.
\end{equation}
Imposing CP: $\{ \Delta {\tilde{m}}^2_{ij} = \Delta {\tilde{\bar m}}^2_{ij},\,
{\tilde J}_{\alpha\beta}^{ij} = \tilde{\bar J}_{\alpha\beta}^{ij}\}$
leads to $\barasym{T}_{\alpha\beta} = \asym{T}_{\alpha\beta}$.
Conversely, imposing CPT: $\{ \Delta {\tilde{m}}^2_{ij} = \Delta {\tilde{\bar m}}^2_{ij},\,
{\tilde J}_{\alpha\beta}^{ij} = (\tilde{\bar J}_{\alpha\beta}^{ij})^* \}$, 
as happens in vacuum, leads to 
$\barasym{T}_{\alpha\beta} = -\asym{T}_{\alpha\beta}$.
Therefore, we expect from symmetry principles that all terms in $\asym{T}_{\alpha\beta}$
will also appear in $\barasym{T}_{\alpha\beta}$,
either with the same sign ---and so conserving CP and violating CPT---
or with opposite sign ---violating CP, and CPT invariant.



A consistent separation of the T asymmetry
into two disentangled components,
$\asym{T}_{\alpha\beta} = \comp{T}{CP}_{\alpha\beta}+\comp{T}{CPT}_{\alpha\beta}$,
implies the decomposition of the antineutrino asymmetry,
in terms of the same components as $\asym{T}_{\alpha\beta}$, into
$\barasym{T}_{\alpha\beta} = -\comp{T}{CP}_{\alpha\beta}+\comp{T}{CPT}_{\alpha\beta}$.
Explicit expressions for these two components can be read 
from Eqs.~(\ref{eq:asymT}, \ref{eq:asymTbar}) as
\begin{subequations}
	\label{eqs:compsAT}
	\begin{align}
		\label{eq:AT(CP)}
	    \comp{T}{CP}_{\alpha\beta} &= 
	    -2\sum_{j<i} \left[
	   	\mathrm{Im}~\tilde J_{\alpha\beta}^{ij}\,\sin 2\tilde \Delta_{ij}
		-\mathrm{Im}~\tilde {\bar J}_{\alpha\beta}^{ij}\,\sin 2\tilde {\bar \Delta}_{ij}
		\right]\,,\\
		\label{eq:AT(CPT)}
	    \comp{T}{CPT}_{\alpha\beta} &=
	    -2\sum_{j<i} \left[
	   	\mathrm{Im}~\tilde J_{\alpha\beta}^{ij}\,\sin 2\tilde \Delta_{ij}
		+\mathrm{Im}~\tilde {\bar J}_{\alpha\beta}^{ij}\,\sin 2\tilde {\bar \Delta}_{ij}
		\right]\,.
	\end{align}
\end{subequations}

Notice that the CP-odd CPT-invariant component (\ref{eq:AT(CP)}) of the T asymmetry
is the same as the genuine T-odd CPT-invariant component (\ref{eq:ACP(T)}) of the CP asymmetry.
This fact can be understood from the \emph{asymmetry sum rule}\footnote{
	This identity is trivially proved
	from the definition of the three asymmetries
	as differences of oscillation probabilities,
	$P(\nu_\alpha \to \nu_\beta) - P(\bar\nu_\beta \to \bar\nu_\alpha) =
	P(\nu_\alpha \to \nu_\beta) - P(\bar \nu_\alpha \to \bar \nu_\beta) +
	P(\bar \nu_\alpha \to \bar \nu_\beta) - P(\bar\nu_\beta \to \bar\nu_\alpha)
	$.
}
\begin{equation}
	\label{eq:sumrule}
	\asym{CPT}_{\alpha\beta} = \asym{CP}_{\alpha\beta} + \barasym{T}_{\alpha\beta}\,.
\end{equation}
In the CPT-invariant limit, this relation fixes the obtained identity
$\comp{T}{CP}_{\alpha\beta} = \comp{CP}{T}_{\alpha\beta}$.
The fact that these components are always CPT-invariant
ensures this equality is valid even if there is CPT violation. 

On the other hand, a new component 
$\comp{T}{CPT}_{\alpha\beta}$ in (\ref{eq:AT(CPT)})
appears in the T asymmetry not seen in the CP asymmetry. 
Contrary to the matter-induced $\comp{CP}{CPT}_{\alpha\beta}$~(\ref{eq:ACP(CPT)}),
a non-vanishing value of this component requires
the combined effect of both genuine CP violation \emph{and}
the matter-induced CPT violation. Hence it is clear that, in general,
a non-vanishing T asymmetry in matter will present contributions from this CP-even component.

As seen, only the first component 
$\comp{CP}{T}_{\alpha\beta} = \comp{T}{CP}_{\alpha\beta}$
is genuine CPT-invariant as in vacuum, 
whereas the other two,
$\comp{CP}{CPT}_{\alpha\beta}$ and $\comp{T}{CPT}_{\alpha\beta}$,
are CPT-odd as a matter effect.


\section{Observables in terms of vacuum parameters}
\label{sec:Tanalytic}

In the effective Hamiltonian approach describing the behavior of neutrinos 
propagating in uniform matter, its matrix representation in the flavor basis 
is~\cite{MSW-W,Matter-Hamiltonian-Barger, Matter-Hamiltonian-Kuo, Matter-Hamiltonian-Zaglauer, Matter-Hamiltonian-Krastev, Matter-Hamiltonian-Parke}
\begin{equation}
	\label{eq:H}
	H =
	\frac{1}{2E}
	\left\{
		U
	\mqty[
		m_1^2 &0 &0\\
		0 &m_2^2 &0\\
		0 &0 &m_3^2
	] 
	U^\dagger
	+
	\mqty[
		\,a\, &\,0\, &\,0\,\\
		0 &0 &0\\
		0 &0 &0
	] \right\}
	=\frac{1}{2E}\; \tilde U \tilde M^2 \tilde U^\dagger\,,
\end{equation}
where the only difference between neutrinos and antineutrinos comes from
the sign of
(i) the CPV phase $\delta$ in the unitary PMNS matrix $U$
and (ii) the matter parameter $a = 2VE$, with $V$ the matter potential.

In this case, the functional form of the T asymmetry is necessarily
$\asym{T}_{\alpha\beta} \equiv f(a)\sin\delta$,
which leads to the antineutrino asymmetry
$\barasym{T}_{\alpha\beta} = -f(-a)\sin\delta$.
A power expansion of $f(a)$ around the vacuum $a=0$ clearly shows that 
all even powers of $a$ lead to CPT-invariant CP-odd terms,
whereas all odd powers lead to CPT-odd CP-invariant terms.
As a consequence, we discover that, 
even for a uniform (T-symmetric) matter, 
the answer to the title of this paper is positive through
terms of odd-parity in the matter parameter $a$.

\begin{figure}[t]
	\centering
	\begin{subfigure}[t]{0.485\textwidth}
		\begin{tikzpicture}[line width=1 pt, scale=1.5]
			\node at (0,0)[below left]{\includegraphics[width=\textwidth]{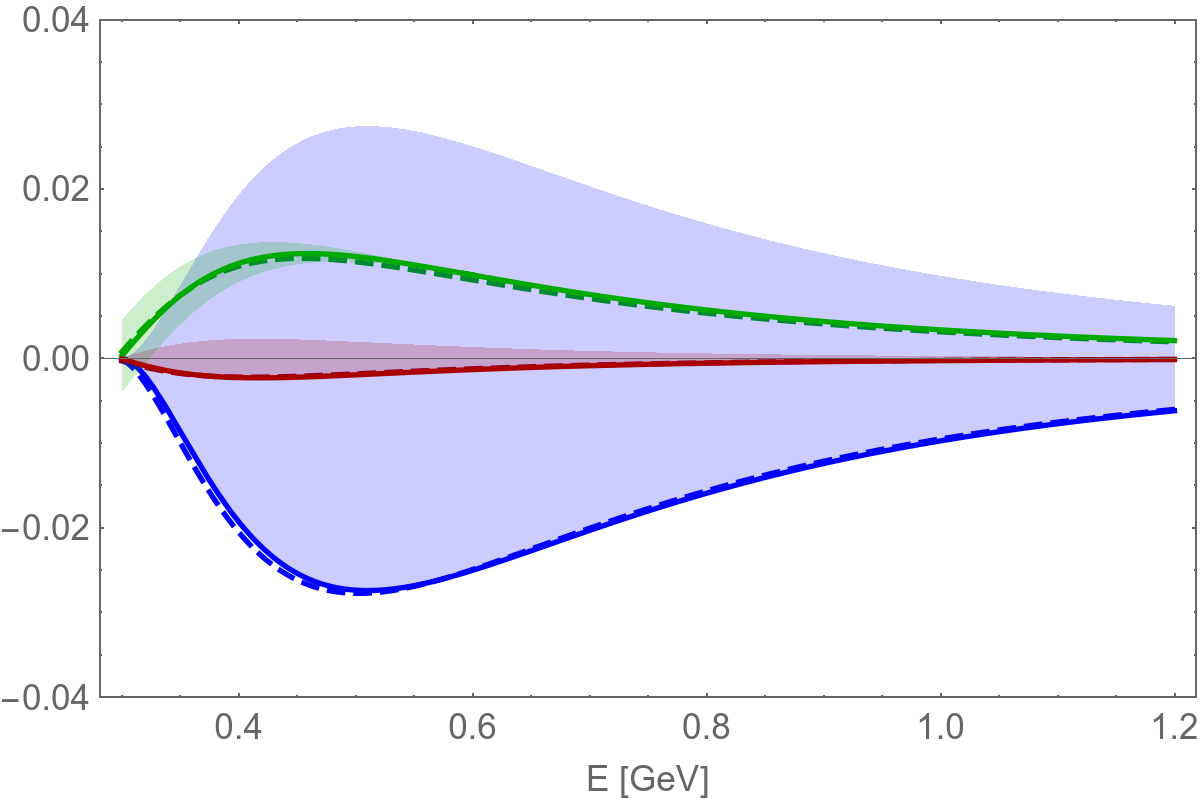}};		
			\node at (-0.1,-0.25)[below left]{$L = 295$~km};
			\node at (-4.0,-0.25)[below left]{NH};
			\node at (-3.5,-1){\tiny\bf \textcolor{ForestGreen}{CP;CPT}};
			\node at (-3.5,-1.8){\tiny\bf \textcolor{BrickRed}{T;CPT}};
			\node at (-3.5,-2.65){\tiny\bf \textcolor{Blue}{T;CP}};
		\end{tikzpicture}
	\end{subfigure}
	\hfill
	\begin{subfigure}[t]{0.485\textwidth}
		\begin{tikzpicture}[line width=1 pt, scale=1.5]
			\node at (0,0)[below left]{\includegraphics[width=\textwidth]{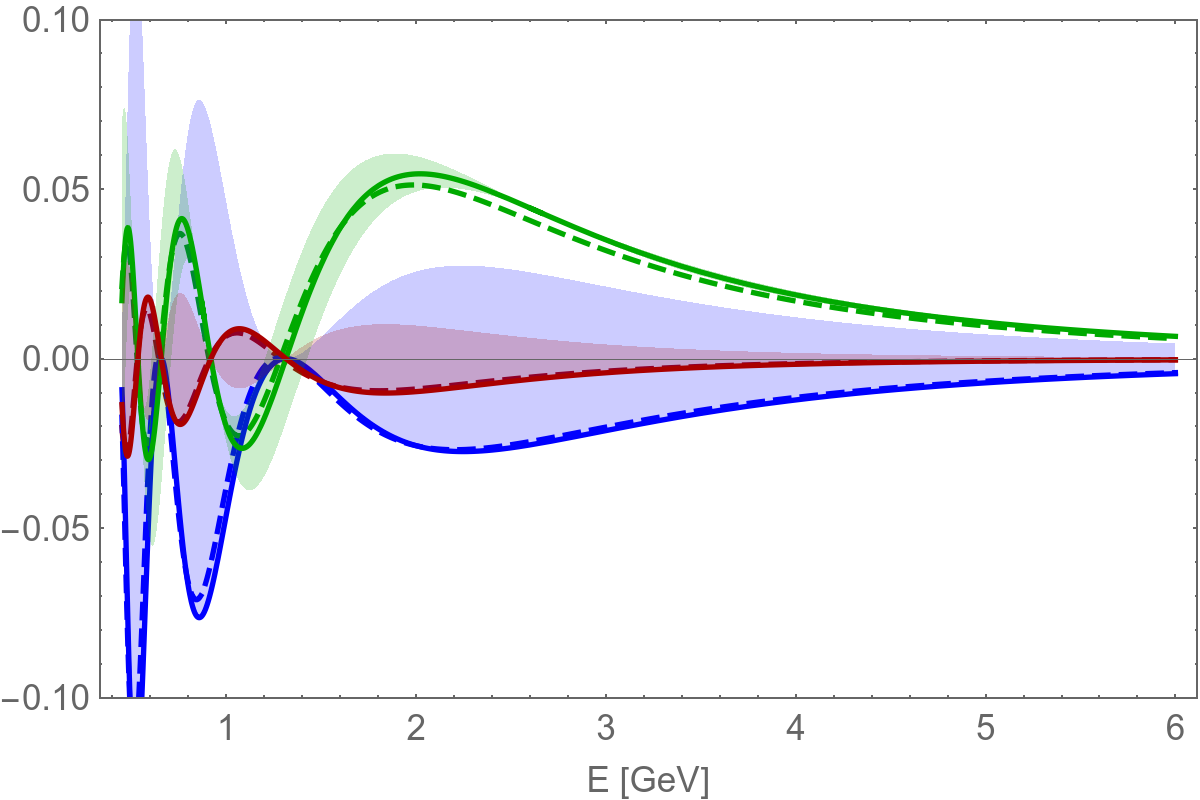}};		
			\node at (-0.1,-0.25)[below left]{$L = 1300$~km};
			\node at (-4.0,-0.25)[below left]{NH};
			\node at (-3.25,-0.65){\tiny\bf \textcolor{ForestGreen}{CP;CPT}};
			\node at (-3.0,-1.8){\tiny\bf \textcolor{BrickRed}{T;CPT}};
			\node at (-3.0,-2.05){\tiny\bf \textcolor{Blue}{T;CP}};
		\end{tikzpicture}
	\end{subfigure}

	\begin{subfigure}[t]{0.485\textwidth}
		\begin{tikzpicture}[line width=1 pt, scale=1.5]
			\node at (0,0)[below left]{\includegraphics[width=\textwidth]{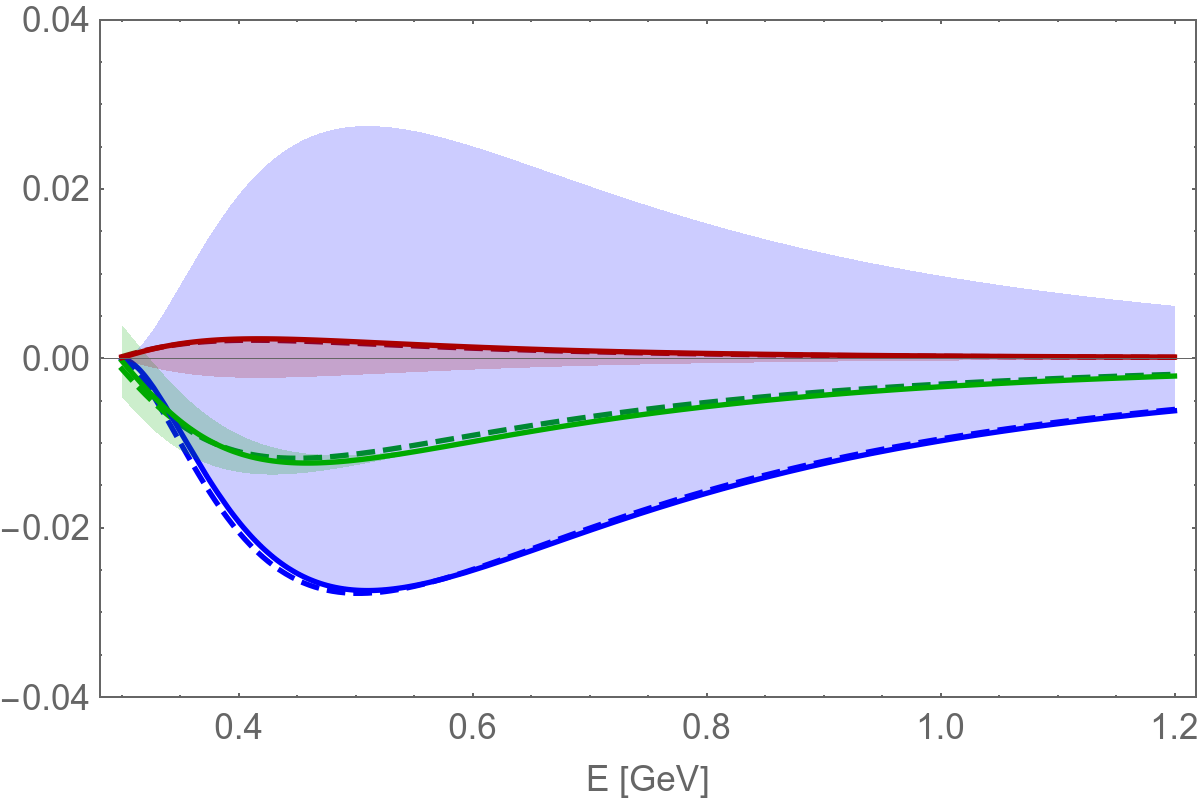}};		
			\node at (-0.1,-0.25)[below left]{$L = 295$~km};
			\node at (-4.0,-0.25)[below left]{IH};
			\node at (-3.5,-2.15){\tiny\bf \textcolor{ForestGreen}{CP;CPT}};
			\node at (-3.5,-1.35){\tiny\bf \textcolor{BrickRed}{T;CPT}};
			\node at (-3.5,-2.65){\tiny\bf \textcolor{Blue}{T;CP}};
		\end{tikzpicture}
	\end{subfigure}
	\hfill
	\begin{subfigure}[t]{0.485\textwidth}
		\begin{tikzpicture}[line width=1 pt, scale=1.5]
			\node at (0,0)[below left]{\includegraphics[width=\textwidth]{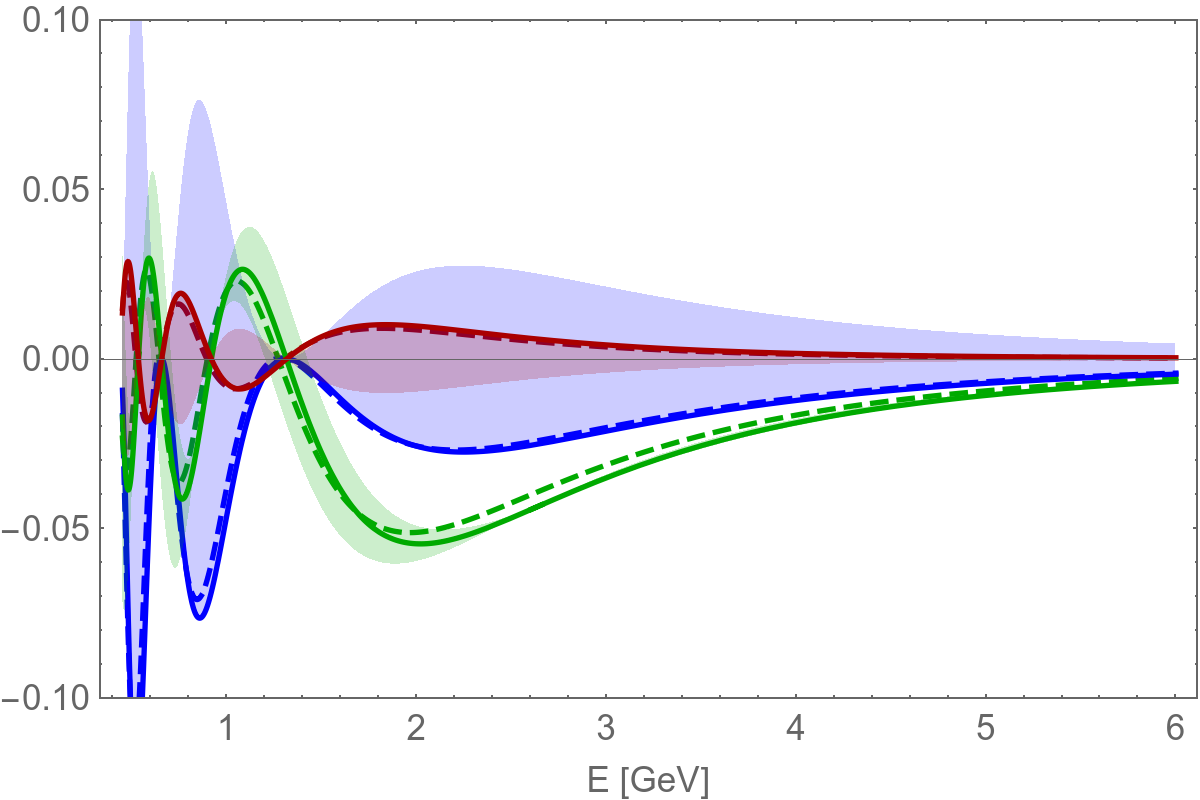}};		
			\node at (-0.1,-0.25)[below left]{$L = 1300$~km};
			\node at (-4.0,-0.25)[below left]{IH};
			\node at (-3.25,-2.5){\tiny\bf \textcolor{ForestGreen}{CP;CPT}};
			\node at (-3.0,-1.35){\tiny\bf \textcolor{BrickRed}{T;CPT}};
			\node at (-3.0,-2.05){\tiny\bf \textcolor{Blue}{T;CP}};
		\end{tikzpicture}
	\end{subfigure}
	\caption{
		Comparison of the CP-odd CPT-invariant (blue) and 
		the CP-invariant CPT-odd (red) components 
		of the $\nu_\mu \leftrightarrow \nu_e$ T asymmetry,
		together with the matter-induced CPT-odd T-invariant (green)
		component of the CP asymmetry. 
		Each label near the lines indicates the two symmetries under which 
		the corresponding asymmetry component is odd.
		Lines show $\comp{T}{CP}_{\mu e}/\sin\delta$, $\comp{T}{CPT}_{\mu e}/\sin\delta$
		and the maximal-CPV ($\cos\delta = 0$) value of $\comp{CP}{CPT}_{\mu e}$,
		from both the exact numerical calculation (dashed) and analytical 
		Eqs.~(\ref{eq:anAT(CP)}, \ref{eq:anAT(CPT)}, \ref{eq:anACP(CPT)}) (solid).
		The shaded regions show all possible values of the components taking into account
		the unknown value of $\sin\delta$.
		Normal Hierarchy assumed in the upper figures,
		Inverted Hierarchy in the lower ones; as seen,
		$\comp{T}{CP}_{\mu e}$ is Hierarchy-independent, 
    	whereas the matter-induced $\comp{T}{CPT}_{\mu e}$ is Hierarchy-odd
		and $\comp{CP}{CPT}_{\mu e}$ is almost~\cite{disentanglingJHEP} Hierarchy-odd.
		The matter-induced CP-even $\comp{T}{CPT}_{\mu e}$ is the same in the
		antineutrino asymmetry case, whereas the other two components change sign.
		Notice the different scales for the two baselines.
	}
	\label{figs:Tcomps}
\end{figure}
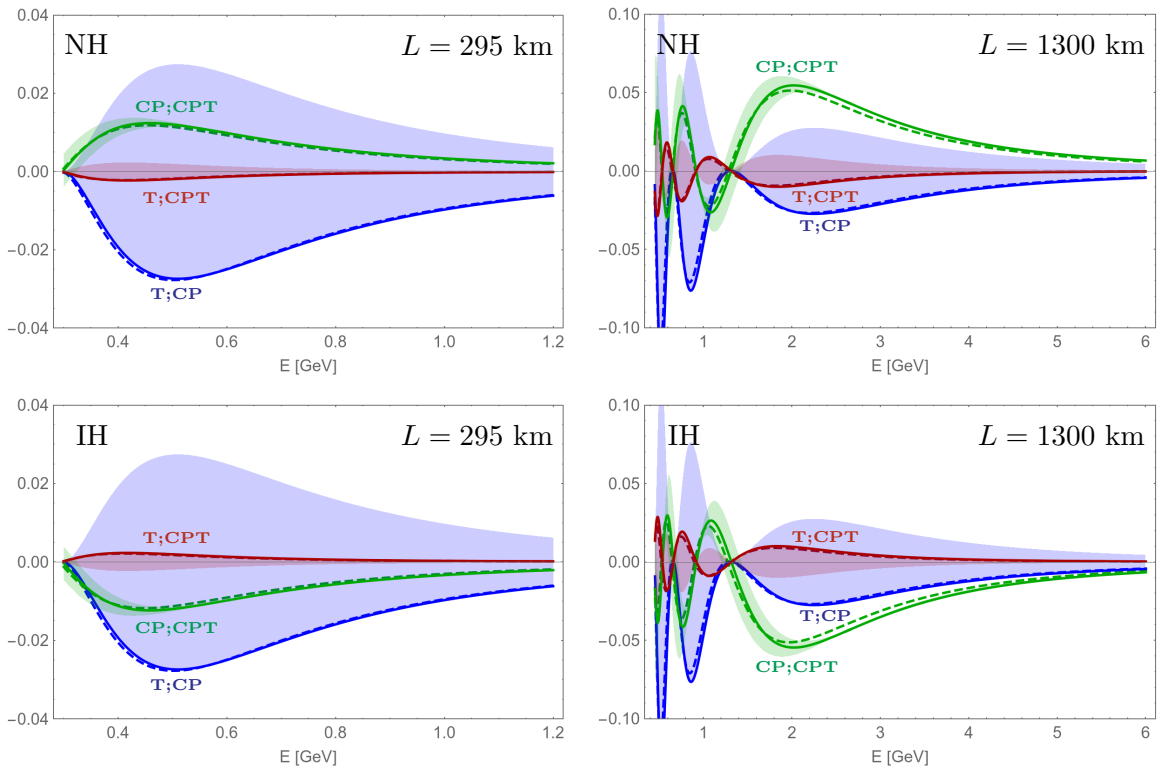

An exact numerical result of the energy dependencies of the two components 
$\comp{T}{CP}_{\mu e}$ and $\comp{T}{CPT}_{\mu e}$
is given in Fig.~\ref{figs:Tcomps} for the two baselines $L = 295,\, 1300$~km.
These baselines correspond to those of the T2HK~\cite{T2HK} and DUNE~\cite{DUNE} experiments,
illustrated here for comparison with the other fake $\comp{CP}{CPT}_{\mu e}$
component of Ref.~\cite{disentanglingJHEP}.
Their direct observability would require terrestrial sources of electron neutrinos
with equal energies to those for muon neutrino sources, not accessible at present.
However, these 
$\asym{T}_{\mu e}$ and $\barasym{T}_{\mu e}$
asymmetries would be natural for a neutrino factory~\cite{NF} and
they are ingredients for an analysis using atmospheric neutrinos.

The vacuum parameters used are those obtained by the global fit in Ref.~\cite{mariam}
\begin{gather}
	\nonumber
	\theta_{12} = 34.5^\circ\,,
	\hspace{1cm}
	\theta_{13} = 8.45^\circ\,,
	\hspace{1cm}
	\theta_{23} = 47.7^\circ\,,\\
	\dm_{21} = 7.5\times 10^{-5}~\text{eV}^2\,,
	\hspace{1cm}
	\dm_{31} = 2.5\times 10^{-3}~\text{eV}^2\,,
\end{gather}
assuming Normal Hierarchy. 
We consider a change of Hierarchy for the same physical configuration, 
which corresponds to compute the Inverted Hierarchy case with the condition
$\dm_{31}|_\mathrm{IH} = -\dm_{32}|_\mathrm{NH}$,
keeping the same values for all other quantities. 
In this way, one is consistent in isolating the observable effects 
due to the change in the ordering of the same neutrino mass states.

\newpage
As expected, the Figure shows that matter effects become relevant at longer baselines.
Since matter effects cannot generate a T asymmetry if there is no genuine CP violation,
both T-odd components vanish when $\sin\delta =0$.
We also find that, as happens in the CP asymmetry, 
there are special configurations at longer baselines 
where the matter-induced term vanishes,
for any value of $\delta$, near a maximal value of the genuine one.

In order to characterize these points analytically,
we consider a perturbative expansion in the small quantities
\begin{gather}
	\label{eq:pert}
	\frac{\dm_{21}}{\dm_{31}} \sim 0.030\,,
	\hspace{0.8cm}
	\abs{U_{e3}} \sim 0.15\,,\\
	\nonumber
	\left(\frac{\dm_{21}}{a}\right)^2 \sim \frac{0.12}{(E/\mathrm{GeV})^2}\,,
	\hspace{0.4cm}
	\left(\frac{a}{\dm_{31}}\right)^2 \sim 0.0083\, (E/\mathrm{GeV})^2\,,
	\hspace{0.4cm}
	\Da^2 
		\sim 0.084\, (L/1000~\mathrm{km})^2\,,
\end{gather}
where $\Da \equiv aL/4E$ is the energy-independent matter-induced phase shift.
Such an expansion is
valid in the energy region
between the two MSW resonances, $\dm_{21} \ll a \ll \abs{\dm_{31}}$.
The relevant ratios of the second line of Eq.~(\ref{eq:pert}) are quadratic
due to the definite $a$-parity of each component of $\asym{T}_{\mu e}$,
so even expanding in them ensures that errors are well kept
at the few percent level.

As thoroughly discussed in Ref.~\cite{disentanglingJHEP}, 
this treatment leads to the disentanglement of the CP asymmetry into
$\asym{CP}_{\mu e} = \comp{CP}{T}_{\mu e} + \comp{CP}{CPT}_{\mu e}$ by
\begin{subequations}
	\begin{align}
		\label{eq:anACP(CPT)}
		\comp{CP}{CPT}_{\mu e} &=
			16\, \Da
			\left[ \frac{\sin\Delta_{31}}{\Delta_{31}}-\cos\Delta_{31} \right]
			\left( S\sin\Delta_{31} + J_r\cos\delta\, \Delta_{21}\cos\Delta_{31} \right)
			+\mathcal{O}(\Da^3)\,,\\
		\label{eq:anACP(T)}
		\comp{CP}{T}_{\mu e} &=
			-16\,J_r \sin\delta\, \Delta_{21}\sin^2\Delta_{31}
			+\mathcal{O}(\Da^2) \,,
	\end{align}
\end{subequations}
where $S \equiv c_{13}^2 s_{13}^2 s_{23}^2$,
$J_r \equiv c_{12} c_{13}^2 c_{23} s_{12} s_{13} s_{23}$,
$\Da\equiv \frac{aL}{4E} \propto L$ and
$\Delta_{ij} \equiv \frac{\dm_{ij} L}{4E} \propto~\!L/E$.
When the analogous treatment is applied to the T asymmetries,
we find that the relevant mixings 
are
\begin{equation}
	\mathrm{Im}~\tilde J_{e \mu}^{12} = 
	- \mathrm{Im}~\tilde J_{e \mu}^{13} = 
	\mathrm{Im}~\tilde J_{e \mu}^{23} = 
		\left(\frac{\dm_{21}}{a} + \frac{\dm_{21}}{\dm_{31}} \right)
		J_r \sin\delta \,,
\end{equation}
and the same for antineutrinos changing the sign of $a$ and $\delta$.
This result is our perturbative approximation of the vacuum-matter invariance
identity~\cite{scott} 
\begin{equation}
	\dmt_{21}\dmt_{31}\dmt_{32}\,
	\mathrm{Im}~\tilde J_{\alpha\beta}^{ij}
	=\dm_{21}\dm_{31}\dm_{32}\,
	\mathrm{Im}~J_{\alpha\beta}^{ij}\,.
\end{equation}
Together with the mass differences
	\begin{alignat}{3}
		\nonumber
		&\dmt_{21} \approx a\,,
		\hspace{1cm}
		&&\dmt_{31} \approx \dm_{31}\,,
		\hspace{1cm}
		&&\dmt_{32} \approx \dm_{31}-a \,,\\
		\label{eq:masses_exp}
		&\dmtbar_{21} \approx \abs{a}\,,
		\hspace{1cm}
		&&\dmtbar_{31} \approx \dm_{31} + \abs{a}\,,
		\hspace{1cm}
		&&\dmtbar_{32} \approx \dm_{31} \,,
	\end{alignat}
we discover that, besides the global $\sin\delta$ factor and
the different $a$-parities, the same combinations of vacuum parameters
as for the CP asymmetry appear. We obtain
%
%
\begin{equation}
	\asym{T}_{\mu e} =
	-16 J_r \sin\delta\, \Delta_{21}\sin\Delta_{31}
	\left[ \sin\Delta_{31} + \Da \left( 
	\frac{\sin\Delta_{31}}{\Delta_{31}}-\cos\Delta_{31}\right) \right]
	+ \mathcal{O}(\Da^2)\,,
\end{equation}
which is disentangled into
\vspace{6pt}\\
\noindent
\boxed{
	\begin{minipage}{0.99\textwidth}
		\vspace{-0.4cm}
		\begin{subequations}
			\label{eqs:Tcomp}
			\begin{align}
				\label{eq:anAT(CP)}
				\comp{T}{CP}_{\mu e} &= 
				-16 J_r \sin\delta\, \Delta_{21}\sin^2\Delta_{31}
				+ \mathcal{O}(\Da^2)\,,\\
				\label{eq:anAT(CPT)}
				\comp{T}{CPT}_{\mu e} &= 
				-16 \Da J_r \sin\delta\, \Delta_{21} \sin\Delta_{31}
				\left[ \frac{\sin\Delta_{31}}{\Delta_{31}}-\cos\Delta_{31} \right]
				+ \mathcal{O}(\Da^3)\,.
			\end{align}
		\end{subequations}
		\vspace{-0.55cm}
	\end{minipage}
}
\vspace{0pt}\\
As expected, the leading term in the genuine component 
in the T asymmetry $\asym{T}_{\mu e}$ 
is its vacuum limit $\comp{T}{CP}_{\mu e}$.
It equals the genuine component $\comp{CP}{T}_{\mu e}$~(\ref{eq:anACP(T)}) of the CP asymmetry,
as required by the discussion in the previous Section.
The leading term of the matter-induced component $\comp{T}{CPT}_{\mu e}$ 
of the T asymmetry explicitly shows 
the combined proportionality to $a \sin\delta$ and,
consequently, it is CP-even.
Whereas $\asym{CP}_{\mu e}$ contains either 
$\sin\delta$ \emph{or} $a$ in its components, 
$\comp{T}{CPT}_{\mu e}$ needs both $\sin\delta$ \emph{and} $a$.

The energy dependence of $\comp{T}{CPT}_{\mu e}$ present in
the oscillating phases is exactly the same as the odd term under
the change of hierarchy in $\comp{CP}{CPT}_{\mu e}$. 
Therefore, it vanishes in the same set of $\delta$-independent zeros
as the matter-induced term $\comp{CP}{CPT}_{\mu e}$ of the CP asymmetry
discussed in Ref.~\cite{disentanglingJHEP}. 
These first-rank zeros correspond to the solutions of 
\begin{equation}
	\label{eq:magicD}
	\tan\Delta_{31} = \Delta_{31}\,,
\end{equation}
whose highest-energy solution happens at
\begin{equation}
	\label{eq:magicE}
	E = 0.92~\mathrm{GeV}\,\frac{L}{1300~\mathrm{km}}
	\,\frac{\abs{\dm_{31}}}{2.5\times 10^{-3}~\mathrm{eV}^2}
\end{equation}
near the second oscillation maximum 
$\Delta_{31} = 3\pi/2$
and near the maximal
---proportional to $\sin\delta$---
genuine $\comp{T}{CP}_{\mu e}$ at
\begin{equation}
	\tan\Delta_{31} = -2\Delta_{31}\,.
\end{equation}

These analytical approximations are compared with the exact numerical results
in Fig.~\ref{figs:Tcomps} for the two chosen baselines of T2HK and DUNE.
We observe that the agreement is excellent.
As is now understood, the T asymmetry is purely genuine 
at the same magic energy (\ref{eq:magicE}) near the second oscillation maximum
at which the experimental CP asymmetry is independent of matter effects. 
Even though a non-vanishing T asymmetry is a proof of genuine CP violation even in matter,
this magic configuration provides a clean matter-independent point
where $\sin\delta$ could be measured
without hierarchy ambiguities.
As shown in Ref.~\cite{disentanglingJHEP}, a modest energy resolution 
would allow the separation of the hierarchy-independent genuine
$\comp{T}{CP}_{\mu e}$ component.

\section{CPT asymmetry lemma}
\label{sec:CPT}

A decomposition of the CPT asymmetry into T-odd CP-invariant and
CP-odd T-invariant components, 
$\asym{CPT}_{\alpha\beta} \equiv P_{\alpha\beta} - \bar P_{\beta\alpha}
= \comp{CPT}{CP}_{\alpha\beta} + \comp{CPT}{T}_{\alpha\beta}$,
is constrained by the CP and T asymmetries decomposition
and the asymmetry sum rule (\ref{eq:sumrule}),
\begin{equation}
	\comp{CPT}{CP}_{\alpha\beta} +
	\comp{CPT}{T}_{\alpha\beta} =
	\comp{CP}{CPT}_{\alpha\beta} +
	\comp{CP}{T}_{\alpha\beta} +
	\comp{T}{CPT}_{\alpha\beta} -
	\comp{T}{CP}_{\alpha\beta}\,.
\end{equation}
This expression implies that, 
due to the separate invariance under the three discrete symmetries,
the following three equalities must be satisfied:
the T-invariant
$\comp{CPT}{CP}_{\alpha\beta} = \comp{CP}{CPT}_{\alpha\beta}$,
the CP-invariant
$\comp{CPT}{T}_{\alpha\beta} = \comp{T}{CPT}_{\alpha\beta}$,
and the CPT-invariant
$\comp{CP}{T}_{\alpha\beta} = \comp{T}{CP}_{\alpha\beta}$.

Therefore,
the decomposition of all asymmetries,
\begin{align}
	\nonumber
	\asym{T}_{\alpha\beta} &= \comp{T}{CPT}_{\alpha\beta} + \comp{T}{CP}_{\alpha\beta}\,,
	\hspace{0.8cm}
	\asym{CPT}_{\alpha\beta} = \comp{CPT}{T}_{\alpha\beta} + \comp{CPT}{CP}_{\alpha\beta}\,,
	\hspace{0.8cm}
	\asym{CP}_{\alpha\beta} = \comp{CP}{T}_{\alpha\beta} + \comp{CP}{CPT}_{\alpha\beta}\,,
	\\
	\label{eq:asymsDecomp}
	\barasym{T}_{\alpha\beta} &= \comp{T}{CPT}_{\alpha\beta} - \comp{T}{CP}_{\alpha\beta}\,,
	\hspace{0.8cm}
	\barasym{CPT}_{\alpha\beta} = \comp{CPT}{T}_{\alpha\beta} - \comp{CPT}{CP}_{\alpha\beta}\,,
\end{align}
is written in terms of only three (per flavor channel) independent components,
since the two superindices of all components commute.
These three independent components correspond to
(i) the genuine component of the CP asymmetry $\comp{CP}{T}_{\alpha\beta}$ (\ref{eq:ACP(T)}),
(ii) the matter-induced component in the CP asymmetry $\comp{CP}{CPT}_{\alpha\beta}$ (\ref{eq:ACP(CPT)})
and (iii) the component $\comp{T}{CPT}_{\alpha\beta}$ (\ref{eq:AT(CPT)}), induced by matter
in presence of genuine CP violation,
breaking the vacuum identity $\barasym{T}_{\alpha\beta} = -\asym{T}_{\alpha\beta}$.
Notice that the interchange of flavor indices
corresponds to a T transformation,
so all T-odd components will be odd under $\alpha \leftrightarrow \beta$,
whereas T-invariant components will remain unchanged.

A calculation of the CPT asymmetry 
following the considerations of the previous Section
results in
\begin{equation}
	\asym{CPT}_{\mu e} =
	16 \Da \left[ \frac{\sin\Delta_{31}}{\Delta_{31}} - \cos\Delta_{31} \right]
	\left[ \Delta_{21} J_r \cos(\delta+\Delta_{31}) + S \sin\Delta_{31}\right]
	+ \mathcal{O}(\Da^3)\,,
\end{equation}
which is consistent with the decomposition 
$\asym{CPT}_{\alpha\beta} = \comp{CPT}{CP}_{\alpha\beta} + \comp{CPT}{T}_{\alpha\beta}$
and the two components 
$\comp{CPT}{CP}_{\mu e} = \comp{CP}{CPT}_{\mu e}$~(\ref{eq:anACP(CPT)})
and $\comp{CPT}{T}_{\mu e} = \comp{T}{CPT}_{\mu e}$~(\ref{eq:anAT(CPT)}),
the first one even in both $L$ and $\sin\delta$,
the second one odd.

Notice that the whole CPT asymmetry is odd in $a$, as expected,
and vanishes at the points $\tan\Delta_{31} = \Delta_{31}$.
This vanishing condition is consistent with previous analytical studies on the
CPT asymmetry~\cite{shunCPT}, where the $(L,\, E)$ values in which $\asym{CPT}_{\mu e} =0$ 
were suggested as a clean configuration on which a non-vanishing 
CPT asymmetry is a proof of intrinsic CPT violation.
It is also consistent with the exact calculation~\cite{BBB} of the
matter-induced CPT and CP asymmetry in the limit $\dm_{21}=0$.
We extend this result to the matter-induced components of the CP and T asymmetries:
the disentanglement~(\ref{eq:asymsDecomp}) of all three asymmetries shows that
$\asym{CPT}_{\mu e}$ vanishes at $\tan\Delta_{31} = \Delta_{31}$
because \emph{both} CPT-odd components separately vanish.
This ensures that 
$\asym{CP}_{\mu e}$ and $\asym{T}_{\mu e}$ probe genuine CP and T violation, respectively,
free from matter effects.
This last statement cannot be ensured only by finding a vanishing CPT asymmetry,
since a configuration in which the two CPT-odd components cancel each other in
$\asym{CPT}_{\mu e} = \comp{CPT}{CP}_{\mu e} + \comp{CPT}{T}_{\mu e}$
would still present fake terms in the CP and T asymmetries.

This discussion clearly shows that,
for neutrinos traveling through the Earth mantle
at energies between the two MSW resonances, $\dm_{21} \ll a \ll \abs{\dm_{31}}$,
there are magic $L/E$ values that satisfy $\tan\Delta_{31} = \Delta_{31}$.
At these points, of which $L/E = 1420$~km/GeV is the highest-energy one,
all three asymmetries probe genuine violation of its associated discrete symmetry.

\section{Conclusions}
\label{sec:conclusions}

This paper represents the culmination of the study of the separate
physics involved in the genuine and fake matter-induced effects for
all observable CP, T and CPT asymmetries in neutrino oscillations.
The starting point was the proof of Disentanglement Theorems allowing
each asymmetry to be written as the sum of two components with definite
transformation properties under the other two symmetries
\begin{equation}
	\asym{CP}_{\alpha\beta} = \comp{CP}{T}_{\alpha\beta} + \comp{CP}{CPT}_{\alpha\beta}\,,
	\quad
	\quad
	\asym{T}_{\alpha\beta} = \comp{T}{CP}_{\alpha\beta} + \comp{T}{CPT}_{\alpha\beta}\,,
	\quad
	\quad
	\asym{CPT}_{\alpha\beta} = \comp{CPT}{CP}_{\alpha\beta} + \comp{CPT}{T}_{\alpha\beta}\,.
\end{equation}
Due to the proved symmetric character of superindices in the components,
only three of them are independent. 
In going from neutrino to antineutrino asymmetries, 
the components with CP superindex change sign, whereas the others remain invariant.
In addition, T-invariant components are symmetric under exchange of flavor indices,
whereas T-odd ones change sign under $\alpha \leftrightarrow \beta$.
The genuine CP, and T, asymmetry is given by the $\comp{CP}{T}_{\alpha\beta}$
component, which is CPT-invariant as in vacuum; 
the other two independent components, 
$\comp{CP}{CPT}_{\alpha\beta}$ and $\comp{T}{CPT}_{\alpha\beta}$, 
are induced by matter in the neutrino propagation and thus they are odd in the matter potential. 
However, there is a very interesting distinction between these two fake components: 
whereas $\comp{CP}{CPT}_{\alpha\beta}$ can be induced by matter alone without
any fundamental CP violation for neutrinos, this paper has demonstrated
a positive answer to the title question on the existence of the matter-induced 
$\comp{T}{CPT}_{\alpha\beta}$ component,
even for T-symmetric matter.
The possible surprise should disappear when realizing that 
the quantum logic is not disjunctive ---either free or matter---
since interference terms are also relevant.
In fact, the analytic solution~(\ref{eq:anAT(CPT)}) for the fake 
$\comp{T}{CPT}_{\mu e}$ component is proportional to both $a$ and $\sin\delta$.

Using the new T Asymmetry Disentanglement~(\ref{eqs:compsAT}), 
the two observable T asymmetries for neutrinos and antineutrinos allow an experimental 
separation of genuine and fake effects under the same conditions of baseline and energy,
\begin{equation}
	\label{eq:concl1}
	\comp{T}{CP}_{\alpha\beta} = \frac{1}{2}\left(
						\asym{T}_{\alpha\beta}-\barasym{T}_{\alpha\beta} \right)\,,
	\quad
	\quad
	\comp{T}{CPT}_{\alpha\beta} = \frac{1}{2}\left(
						\asym{T}_{\alpha\beta}+\barasym{T}_{\alpha\beta} \right)\,,	
\end{equation}
the first component being odd in $L$ and $\sin\delta$, even in $a$;
the second component being odd in $L$, $\sin\delta$ and $a$. 
The genuine component $\comp{T}{CP}_{\mu e}$ is blind to the neutrino mass hierarchy
and the fake component $\comp{T}{CPT}_{\mu e}$ is odd under the change of hierarchy.
The other independent fake component $\comp{CP}{CPT}_{\mu e}$ can then be obtained from the
CP Asymmetry Disentanglement~(\ref{eqs:compsACP}) 
---or, equivalently, from the CPT asymmetry sum rule~(\ref{eq:sumrule})--- as
\begin{equation}
	\label{eq:concl2}
	\comp{CP}{CPT}_{\alpha\beta} = \asym{CP}_{\alpha\beta} - \comp{CP}{T}_{\alpha\beta}\,,
\end{equation}
being even in $L$ and $\sin\delta$ and odd in $a$.

It is now clear that the CP asymmetry in matter 
cannot be written as the T asymmetry in matter plus a CP-odd fake term, 
since matter induces a CP-invariant term in the T asymmetry.
Our matter-induced CP-odd component (\ref{eq:concl2}) 
quantifies the fake effects in the CP asymmetry with the appropriate behavior under T and CPT.

A remarkable result from our precise enough analytic calculation
of the three components in terms of the vacuum parameters is that
the two fake components $\comp{CP}{CPT}_{\mu e}$ and $\comp{T}{CPT}_{\mu e}$ vanish, 
with first-rank zeros, at the same set of \emph{magic $L/E$ values} satisfying
$
	\tan\Delta_{31} = \Delta_{31}\,,
$
independent of the hierarchy, $\delta$ and $a$: only $\abs{\dm_{31}}$ matters.
These solutions appear near the values in which the genuine component 
$\comp{CP}{T}_{\mu e}$ is maximal,
$
	\tan\Delta_{31} = -2\Delta_{31}\,.
$
The joint fulfillment of these two equations 
occurs around the solutions 
for the even 
oscillation maxima in the transition probabilities, $\sin^2\Delta_{31}=1$.


All in all, the conditions for a direct evidence of genuine CP and T violation 
for neutrino oscillations in matter by means of the measurement 
of an observable component odd under the symmetry are now met. 
One may proceed by 
(a) measuring the genuine CP or T asymmetry at the magic energies where matter effects vanish, 
or (b) follow the program outlined in Eq.~(\ref{eq:concl1}) 
at any chosen $(L;\, E)$ configuration to extract it. 
We emphasize that this genuine component is independent of the hierarchy. 
So global fits in which the resulting $\delta$ phase is dependent on the assumed hierarchy 
say that true CP violating terms, 
proportional to $\sin\delta$ and even in $a$, play a minor role in the intervening observables. 
On the contrary, the two fake components have clear sign information on the hierarchy, 
being odd in $a$ and containing $\delta$-independent, $\cos\delta$ and $\sin\delta$ terms. 
The relative signs of genuine $\comp{T}{CP}_{\mu e}$ and fake $\comp{T}{CPT}_{\mu e}$ components 
are equal for normal hierarchy, 
the last one being smaller in magnitude at all energies and baselines. 
This is in contrast with the fake $\comp{CP}{CPT}_{\mu e}$ component, 
which is larger in magnitude than the genuine $\comp{CP}{T}_{\mu e}$ at large energies for long baselines.

\section*{Acknowledgments}

This research has been supported by MINECO Project FPA 2017-84543-P, 
Generalitat Valenciana Project GV PROMETEO 2017-033 and 
Severo Ochoa Excellence Centre Project SEV 2014-0398. 
A.S. acknowledges the MECD support through the FPU14/04678 grant,
and he is indebted to the hospitality of the Durham IPPP, where this work was finished.


\small
\linespread{1}\selectfont

\end{document}